# Data assimilation of flow MRI data into RANS models with algebraic closures

**Claire Namuroy[1], Matthew P. Juniper[1], Priya Nair[2], Michael Loecher[3], Daniel B. Ennis[4], Alison Marsden[2] and Alexandros Kontogiannis[1]**

[1]Department of Engineering, Trumpington Street, Cambridge CB2 1PZ, UK

[2]Department of Bioengineering, Stanford University, Stanford, CA, USA

[3]Department of Radiology, Stanford University, Stanford, CA, USA

[4]Stanford Cardiovascular Institute, Stanford University, Stanford, CA, USA

**Corresponding author:** Matthew P. Juniper, mpj1001@cam.ac.uk

We adopt the Bayesian inference framework to solve an inverse Reynolds-Averaged Navier–Stokes (RANS) problem for the approximate posterior probability distribution of the turbulence model parameters and inlet boundary conditions of a confined turbulent jet. The data are noisy 3D flow MRI measurements of a Newtonian fluid flowing through the Food and Drug Administration (FDA) nozzle geometry. We assimilate this into RANS using two algebraic turbulence models based on the mean shear rate magnitude and the turbulence kinetic energy. We demonstrate that the inferred models are able to reconstruct the measured mean flow velocities without overfitting and we provide uncertainty estimates for the model parameters. The methodology can readily be extended to more complex RANS models, provided that they remain differentiable.

## 1. Introduction

Flow MRI is a non-invasive experimental technique that provides time-resolved flow velocity measurements in 3D space. Despite its advantages, flow MRI suffers from low spatiotemporal resolution and imaging artefacts, which prevent the turbulent scales of motion from being directly captured. Computational Fluid Dynamics (CFD) provides a means of simulating these scales and enables flow characterisations devoid of noise and artefacts, although they may contain model error. Most CFD studies of biological flows assume laminar flow or use standard Reynolds-Averaged Navier–Stokes (RANS) or Large Eddy Simulation (LES) closures from the literature rather than Direct Numerical Simulation (DNS), which is too computationally expensive. These models, which have been tuned predominantly for aerospace applications, have been shown to have limited accuracy unless informed by experimental data (Lui et al. 2020). Furthermore, CFD modelling of cardiovascular flows relies on the accurate specification of the geometry and boundary conditions, which are not known *a priori*.

To address these challenges, the reconstruction of the mean flow can be treated as an inverse Reynolds-Averaged Navier–Stokes (RANS) problem with unknown parameters. In the wider turbulence community, Foures et al. (2014) closed the RANS equations by inferring a source term in the momentum equation from time-averaged DNS data of a 2D cylinder flow. Foures' variational approach was also used by Singh et al. (2017) and Franceschini et al. (2020), who tuned additive and multiplicative corrections in the Spalart-Allmaras transport equation to adjust the eddy viscosity and the level of turbulence production. Although the inferred models generally fitted the data well, it was shown that this corrective momentum forcing approach led to ill-posed reconstructions owing to the lack of mathematical constraints on the source term (Franceschini et al. 2020). The use of a multiplicative correction as a control parameter was more robust, but less accurate.

An alternative approach consists in superresolving the velocimetry data. Recently, Villié et al. (2025) trained two physics-informed neural networks (PINNs) to assimilate experimental measurements of a transitional stenotic flow and correct for imaging artefacts. Unlike adjoint-based methods, which solve a saddle-point problem, PINNs enforce the flow physics as soft constraints by minimising a weighted residual loss. This means that PINNs seldom satisfy the physical constraints and exhibit strong sensitivity to observational noise.

Across all these studies, data assimilation was performed in a deterministic setting, in that the uncertainty in the parameter estimation introduced by modelling assumptions and experimental error was not quantified. Bayesian inference, on the other hand, is a probabilistic technique that allows the modeller to encode prior physical knowledge and to estimate the expected values and uncertainties of the model parameters from experimental observations. Kontogiannis et al. (2025a) showed that the Bayesian framework could be used to jointly reconstruct noisy velocity fields and learn the most likely rheological parameters of a Carreau shear-thinning model. The inferred model was validated against a separate rheometry experiment. The Bayesian inverse methodology was also applied to a turbulent flow MRI experiment through the Food and Drug Administration's (FDA) benchmark nozzle (Kontogiannis et al. 2024). In the current paper, we build on the work of Kontogiannis et al. (2024) to solve an inverse RANS problem for the effective viscosity field and the inlet boundary conditions of a confined turbulent jet at a Reynolds number of 6500. We consider two algebraic closures, based on the mean shear rate magnitude and measurements of the turbulence kinetic energy (TKE), for their mathematical simplicity and their differentiability properties. We infer the most likely parameters of each model and rigorously quantify their posterior uncertainties. To the best of the authors' knowledge, no other studies have used TKE data from flow MRI to reconstruct turbulent mean flows.

## 2. Bayesian inversion of the Reynolds-Averaged Navier–Stokes problem

We reconstruct the flow field by solving a *Bayesian inverse RANS problem* (Kontogiannis et al. 2024). Formally, we assume that the flow MRI data, $\boldsymbol{u}^{\star}$, are well-described by an algebraic turbulence model with parameters $\boldsymbol{x}^{\circ}$. The data-model discrepancy $\boldsymbol{\epsilon}$ is a Gaussian white noise given by

$$\boldsymbol{u}^{\star} - \mathcal{Z}\boldsymbol{x}^{\circ} = \boldsymbol{\epsilon} \sim \mathcal{N}(\mathbf{0}, \mathcal{C}_{\boldsymbol{u}^{\star}}) \tag{2.1}$$

where the nonlinear operator $\mathcal{Z}$ maps parameters from the parameter space to solutions $\boldsymbol{u}^{\circ}$ projected to the data space. The data $\boldsymbol{u}^{\star}$ consist of velocimetry measurements. An additional set of turbulent kinetic energy measurements, $k^{\star}$, is obtained from intravoxel standard deviation of the MRI signal. In this paper, we supply $k^{\star}$ as an input to the $k$-based algebraic model, but assimilate only $\boldsymbol{u}^{\star}$. The assimilation of TKE data requires the specification of a transport equation for $k$, and will be the subject of future work. The model parameters are assumed to follow a Gaussian distribution $\mathcal{N}(\bar{\boldsymbol{x}}, C_{\bar{\boldsymbol{x}}})$, where $\bar{\boldsymbol{x}}$ represents the prior expected value and $C_{\bar{\boldsymbol{x}}}$ is the prior covariance operator. To assimilate the flow MRI measurements, we apply Bayes' theorem, which provides an expression for the posterior probability density function (p.d.f.) of the model parameters given the data

$$\pi(\boldsymbol{x}|\boldsymbol{u}^\star) \propto \pi(\boldsymbol{u}^\star|\boldsymbol{x})\pi(\boldsymbol{x}) = \exp(-\frac{1}{2}||\boldsymbol{u}^\star - \mathcal{Z}\boldsymbol{x}||^2_{C_{\boldsymbol{u}^\star}} - \frac{1}{2}||\boldsymbol{x} - \bar{\boldsymbol{x}}||^2_{C_{\bar{\boldsymbol{x}}}}) \tag{2.2}$$

where $\pi(\boldsymbol{x})$ is the prior p.d.f. of $\boldsymbol{x}$, $\pi(\boldsymbol{u}^\star|\boldsymbol{x})$ is the likelihood function and $||\cdot,\cdot||^2_C = \langle\cdot, C^{-1}\cdot\rangle$ is the covariance-weighted $L^2$-norm. The most likely parameters, $\boldsymbol{x}^\circ$, maximise the posterior p.d.f. and are obtained as the solution of the minimisation problem

$$\boldsymbol{x}^\circ \equiv \underset{\boldsymbol{x}}{\operatorname{argmin}}\, \mathscr{J}, \quad \text{where} \quad \mathscr{J} = \underbrace{\frac{1}{2}||\boldsymbol{u}^\star - \mathcal{Z}\boldsymbol{x}||^2_{C_{\boldsymbol{u}^\star}}}_{\text{data-model discrepancy}} + \underbrace{\frac{1}{2}||\boldsymbol{x} - \bar{\boldsymbol{x}}||^2_{C_{\bar{\boldsymbol{x}}}}}_{\text{prior term}} \tag{2.3}$$

The *maximum-a-posteriori* (MAP) estimation problem (2.3) is solved iteratively using a damped Broyden–Fletcher–Goldfarb–Shanno (BFGS) method (Goldfarb et al. 2020). Writing the first-order Taylor series approximation of $\mathcal{Z}$ around the $k^{\text{th}}$ iterate as

$$\mathcal{Z}\boldsymbol{x} \simeq \mathcal{Z}\boldsymbol{x}_k + \mathcal{G}_k(\boldsymbol{x} - \boldsymbol{x}_k) \tag{2.4}$$

and enforcing the first-order conditions of optimality on problem (2.3) yields the following update rule (Tarantola 2005)

$$\boldsymbol{x}_{k+1} \leftarrow \boldsymbol{x}_k - \tau_k C_{\boldsymbol{x}_k}(D_{\boldsymbol{x}}\mathscr{J})_k, \tag{2.5}$$

The gradient $(D_{\boldsymbol{x}}\mathscr{J})_k$ is given by

$$(D_{\boldsymbol{x}}\mathscr{J})_k \coloneqq -\mathcal{G}_k^* C^{-1}_{\boldsymbol{u}^\star}(\boldsymbol{u}^\star - \mathcal{Z}\boldsymbol{x}_k) + C^{-1}_{\bar{\boldsymbol{x}}}(\boldsymbol{x} - \boldsymbol{x}_k) \tag{2.6}$$

where $\mathbb{R} \ni \tau_k > 0$ is the step size determined by a line-search procedure at the $k^{\text{th}}$ iteration and $\mathcal{G}_k^*$ is the adjoint of $\mathcal{G}_k$. The posterior covariance operator $C_{\boldsymbol{x}_k}$ at the $k^{\text{th}}$ iteration is obtained from the inverse of the Hessian matrix

$$C_{\boldsymbol{x}_k} \coloneqq (\mathcal{G}_k^* C^{-1}_{\boldsymbol{u}^\star}\mathcal{G}_k + C^{-1}_{\bar{\boldsymbol{x}}})^{-1} \tag{2.7}$$

### 2.1. *The generalised RANS problem*

The forward operator of the 3D RANS boundary value problem, $\mathcal{Z}$, takes the form

$$\begin{gathered}\boldsymbol{u}\cdot\nabla\boldsymbol{u} - \nabla\cdot(2\nu_e\nabla^s\boldsymbol{u}) + \nabla p = \boldsymbol{0} \quad \text{and} \quad \nabla\cdot\boldsymbol{u} = 0 \quad \text{in} \quad \Omega, \\ \boldsymbol{u} = \boldsymbol{0} \quad \text{on} \quad \Gamma, \quad \boldsymbol{u} = \boldsymbol{g}_i \quad \text{on} \quad \Gamma_i, \quad -2\nu_e\nabla^s\boldsymbol{u} + p\boldsymbol{\nu} = \boldsymbol{g}_o \quad \text{on} \quad \Gamma_o\end{gathered} \tag{2.8}$$

where $\boldsymbol{u}$ is the mean flow velocity, $p \leftarrow p/\rho + \frac{2}{3}k$ is the mean modified pressure, $\rho$ is the density and $\nu_e \coloneqq \nu_l + \nu_t$ is the effective kinematic viscosity, written in terms of its laminar and turbulent components. The mean flow strain-rate tensor is defined as $(\nabla^s\boldsymbol{u})_{ij} \coloneqq \frac{1}{2}(\partial_i u_j + \partial_j u_i)$. The Dirichlet boundary condition (b.c.) at the inlet $\Gamma_i$ and the Neumann b.c. at the outlet $\Gamma_o$ are denoted by $\boldsymbol{g}_i$ and $\boldsymbol{g}_o$, respectively. $\Gamma$ is the no-slip wall.

The Reynolds stresses are absorbed into the eddy viscosity, $\nu_t$, under the Boussinesq hypothesis. We model $\nu_t$ algebraically using the compound mixing length model (Kontogiannis et al. 2024)

$$l_m = \alpha\mathcal{H}_\eta(d_s; c)\mathcal{F}_\epsilon(d_s - d_{s_0}) + \beta(1 - \mathcal{H}_\eta(d_s; c))d_w \tag{2.9}$$

where $\alpha, \beta \in \mathbb{R}$ are constants and $d_{s_0} \in \mathbb{R}$ is an offset. $\mathcal{H}_\eta$ and $\mathcal{F}_\epsilon$ are smooth activation functions, defined as

$$\mathcal{H}_\eta(d_s; c) := \left(1 + \exp\left[\frac{d_s^2 - c^2}{\eta c^2}\right]\right)^{-1} \tag{2.10}$$

$$\mathcal{F}_\epsilon(d_s - d_{s_0}) := \frac{1}{2}\left(d_s - d_{s_0} + \sqrt{(d_s - d_{s_0})^2 + \epsilon}\right) \approx \max(d_s - d_{s_0}, 0) \tag{2.11}$$

with $\eta = 0.2$ and $\epsilon = 1 \times 10^{-8}$. When $\mathcal{H}_\eta \rightarrow 1$, the mixing length varies linearly with streamwise distance, $d_s$. As $\mathcal{H}_\eta \rightarrow 0$, $l_m$ scales instead with wall-normal distance, $d_w$. This allows the eddy viscosity model to be applied to flows featuring both free-shear and wall-bounded flow regions. The inclusion of $\mathcal{F}_\epsilon$ ensures that the mixing length model (2.9) remains non-negative and differentiable. We consider the Smagorinski and Prandtl-Kolmogorov eddy viscosity relations, defined respectively as $\nu_t := l_m^2 \dot{\gamma}$ and $\nu_t := l_m \sqrt{k}$ (Pope 2000). Here, $\dot{\gamma} := \sqrt{2\nabla^s \boldsymbol{u} : \nabla^s \boldsymbol{u}}$ represents the mean shear rate magnitude. To obtain the smoothed turbulence kinetic energy field, $k$, we convolve the data, $k^\star$, with a normalised exponential covariance kernel, $C_k$, in Fourier space. We define $(C_k)_{ij} := \frac{\pi^{3/2} l^3}{3!\Gamma(5/2)} \exp\left(\frac{-||x_i - x_j||_2}{l}\right)$ where the covariance lengthscale $l = 2h^\star$ and $h^\star = 1$mm is the data resolution. Chapter 2.2.5 of Kontogiannis (2023) contains a detailed discussion of the smoothing procedure.

### 2.2. *Operators $\mathcal{Z}$, $\mathcal{G}$*

The action of $\mathcal{Z} := \mathcal{S}\mathcal{Q}$ can be viewed as a mapping $\mathcal{Q} \colon \boldsymbol{P} \rightarrow \boldsymbol{M}$ from the model parameter space $\boldsymbol{P}$ to the model solution space $\boldsymbol{M}$, followed by a projection $\mathcal{S} \colon \boldsymbol{M} \rightarrow \boldsymbol{D}$ from $\boldsymbol{P}$ to the data space $\boldsymbol{D}$. In a similar vein, we define $\mathcal{G}_k := \mathcal{S}\mathcal{A}_k$, where the linearised operator $\mathcal{A}_k \equiv \left((D_{\boldsymbol{u}}^{\mathscr{M}})^{-1} D_{\boldsymbol{x}}^{\mathscr{M}}\right)_k$ is constructed from the inverse Jacobian of the RANS problem, $(D_{\boldsymbol{u}}^{\mathscr{M}})^{-1}$, and the generalised gradient of the velocity field with respect to the parameters, $D_{\boldsymbol{x}}^{\mathscr{M}}$. The derivation of the operators $\mathcal{S}, \mathcal{Q}$ and $\mathcal{A}$ from the weak form, $\mathscr{M}$, of (2.8) can be found in Kontogiannis et al. (2025b). It follows from these definitions that the contribution of the parameters $\boldsymbol{p}_\mu$ of the viscosity model to $(D_{\boldsymbol{x}}\mathscr{J})_k$ is given by

$$\delta \boldsymbol{p}_\mu := (D_{\boldsymbol{p}_\mu}^{\mathscr{M}})_k^* \left((D_{\boldsymbol{u}}^{\mathscr{M}})^*\right)_k^{-1} \mathcal{S}^* C_{\boldsymbol{u}^\star}^{-1}(\boldsymbol{u}^\star - \mathcal{S}\boldsymbol{u}_k) \tag{2.12}$$

where $D_{\boldsymbol{p}_\mu}^{\mathscr{M}}$ and $D_{\boldsymbol{u}}^{\mathscr{M}}$ are the gradients of $\mathscr{M}$ with respect to the parameters of the effective viscosity model and the velocity field, respectively. To compute $\delta \boldsymbol{p}_\mu$, we solve the adjoint system of equations for the mean flow $\left(D_{\boldsymbol{u}}^{\mathscr{M}}\right)_k^* \boldsymbol{v}_k = \mathcal{S}^* C_{\boldsymbol{u}^\star}^{-1}(\boldsymbol{u}^\star - \mathcal{S}\boldsymbol{u}_k)$, as derived in Kontogiannis et al. (2025b). Using the adjoint velocity, $\boldsymbol{v}_k$, equation (2.12) is rewritten as

$$\delta \boldsymbol{p}_\mu := (D_{\boldsymbol{p}_\mu}^{\mathscr{M}})_k^* \boldsymbol{v}_k = 2\int_\Omega (D_{\boldsymbol{p}_\mu}\mu_e)_k (\nabla^s \boldsymbol{u}_k : \nabla^s \boldsymbol{v}_k) \tag{2.13}$$

where $(D_{\boldsymbol{p}_\mu}\mu_e)_k$ is the gradient of the effective viscosity model with respect to its parameters.

## 3. Flow MRI experiment of a confined turbulent jet

We test the proposed algorithm on *in vitro* imaging-based measurements of a turbulent flow of Newtonian fluid through the FDA benchmark nozzle. The experimental setup is presented in figure 1. The flow enters the nozzle through a conical inlet baffle and is then straightened through a grid of $0.2\text{cm}\times0.2\text{cm}$ square channels with a depth of $0.1\text{cm}$. This ensures the removal of flow disturbances and the creation of reproducible inflow conditions. The flow is subsequently accelerated through a converging duct before entering the throat section with a diameter of $1\text{cm}$. At the entrance to the test chamber, the diameter increases threefold. The flow forms a jet at the entrance, surrounded by an annular recirculation zone. The flow transitions to fully-turbulent pipe flow downstream of this recirculation zone. The mass flow rate is set to $200\text{mL}/\text{s}$, corresponding to a bulk Reynolds number of $Re = 6500$ at the throat. A complete description of the experimental conditions and the MRI imaging protocol is provided in Kontogiannis et al. (2024).

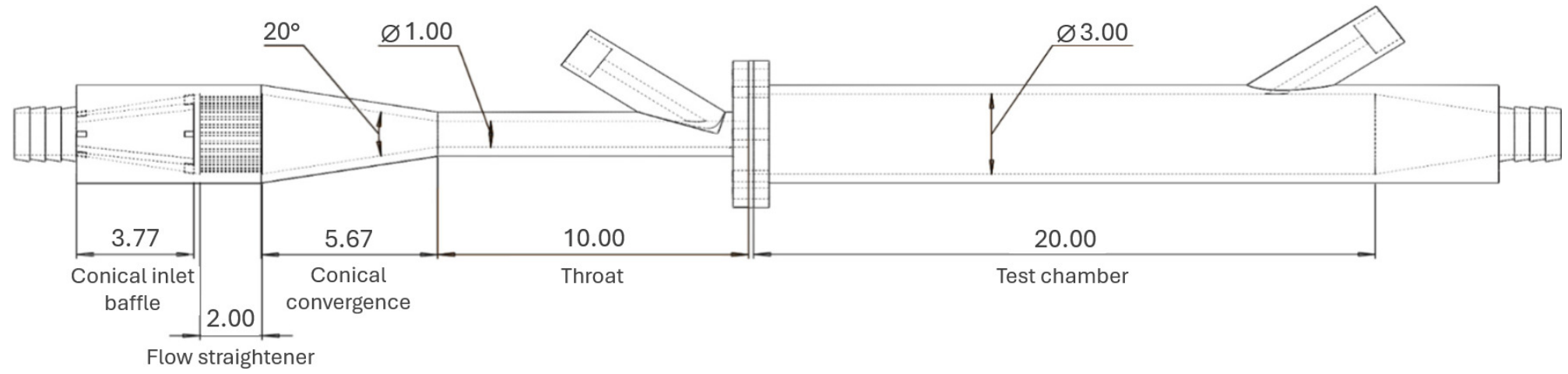

Figure 1. Schematic of the FDA nozzle (adapted from Kontogiannis et al. (2024)). Lengths are in cm.

## 4. Joint flow field reconstruction and algebraic model parameter learning

We use the methodology described in section 2 to learn the most likely effective viscosity parameters of the $\dot{\gamma}$-based and $k$-based models, along with the mean inflow conditions in each case. The 3D flow is reconstructed on a uniform mesh with resolution $h = 0.75\text{mm}$, corresponding to an upsampling factor of $4/3$ relative to the data resolution. Although we assume zero-Neumann conditions at the outlet and fix the geometry, the proposed framework can naturally be extended to data assimilation of outflow boundary conditions and unknown boundaries (Kontogiannis et al. 2025b).

### 4.1. *Flow field reconstruction*

The reconstructed flow fields, $\boldsymbol{u}^{\circ}$, obtained using the $\dot{\gamma}$- and $k$-based models, are vizualised in figure 2 alongside the data, $\boldsymbol{u}^{\star}$. The velocimetry data are assumed to be contaminated by isotropic white noise with standard deviation $\sigma_u = 3.5\text{cm}/\text{s}$. The prior velocity fields are very similar across models, so we present only the prior flow field from the $\dot{\gamma}$-based model. We see from figure 2a that the streamlines poorly resolve the recirculation zones due to the presence of noise and artefacts in the data, while the inferred mean velocity fields in figures 2c and 2d can reconstruct these low-velocity regions. There is good agreement between the jet lengths and peak velocity magnitudes from the inferred models.

From figure 3 we see that the data-model discrepancies are very similar across models after the data is assimilated. Although the posterior reconstructions show a significant reduction in data-model discrepancy relative to the prior assumptions, we observe some residual discrepancies near the inlet and in the shear layers. One source of discrepancy is the model error arising from the simplicity of the algebraic models. To further decrease the posterior discrepancies, the inversion framework could be applied to more complex turbulence models that account for flow history effects. The posterior data-model discrepancies also have a component of unquantified systematic error stemming from the flow imaging technique. Near the inlet, the experimental mass flow rate measured at each streamwise location deviates by up to 19% from the nominal value of 200 mL/s. This suggests the presence of displacement artefacts, which appear more prominently in high-velocity regions (Nishimura et al. 1991). These artefacts are filtered out by the algorithm because we hardwire mass conservation, further demonstrating that the inferred models can reconstruct the flow without overfitting.

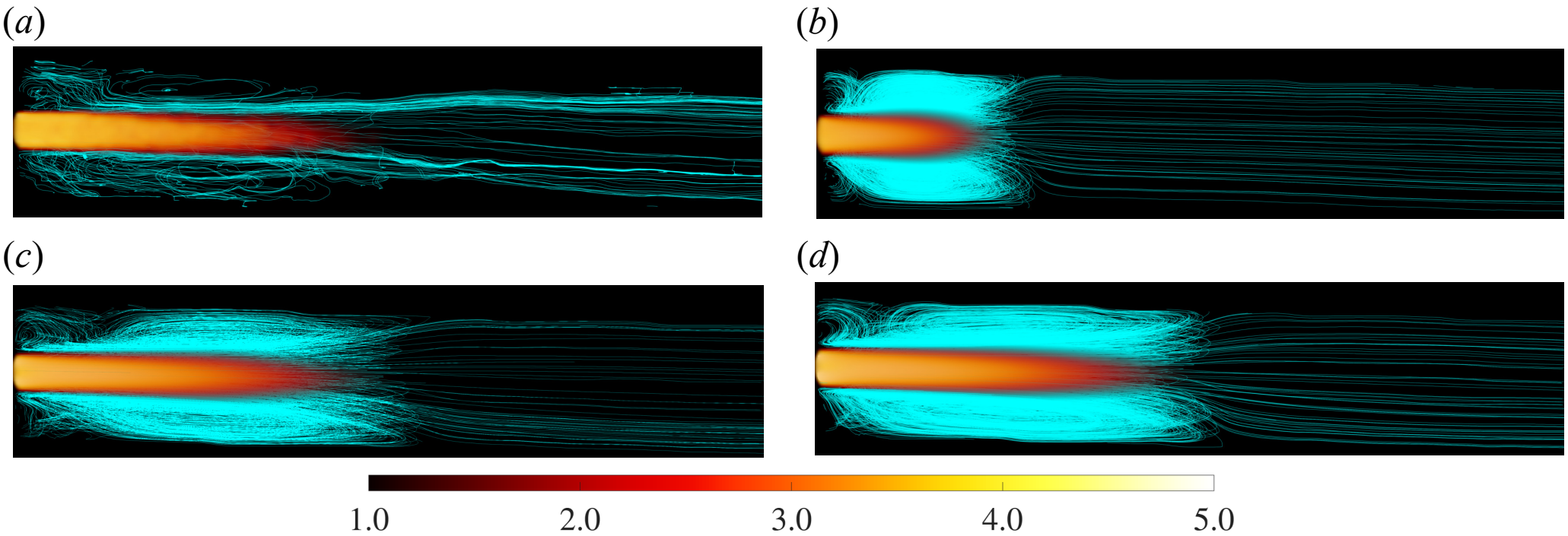

Figure 2. Flow MRI measurements of the velocity magnitude, $\boldsymbol{u}^\star$ (figure 2a), prior (modelled) velocity field, $\bar{\boldsymbol{u}}$, (figure 2b) and RANS model reconstructions, $\boldsymbol{u}^\circ$, obtained with the inferred $\dot{\gamma}$-based model (figure 2c) and $k$-based model (figure 2d). The velocity magnitude shown in figure 2b is computed using the prior $\dot{\gamma}$-based model, and is indistinguishable from the velocity field predicted by the prior $k$-based model. Streamlines of the mean flow are indicated by cyan lines. The colour bar is reported in nondimensional units ($U_{\text{ref}} = 1$ m/s).

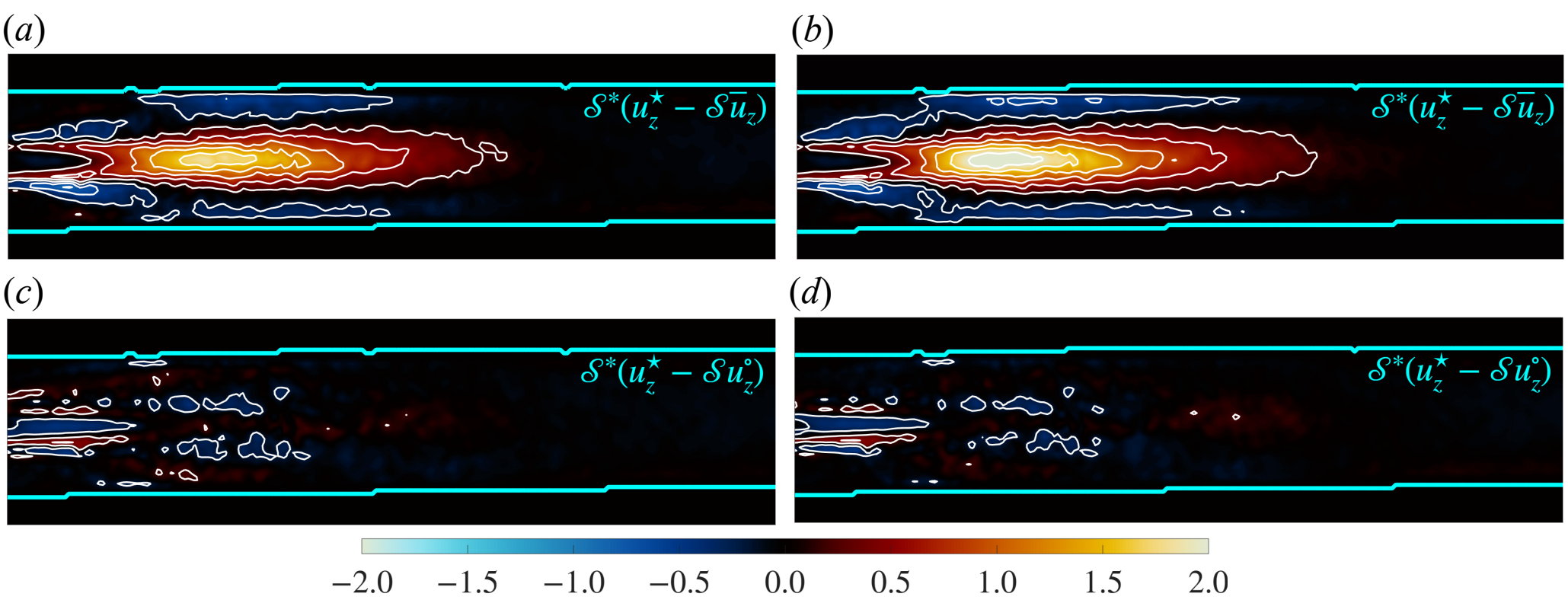

Figure 3. Prior data-model discrepancies, $\mathcal{S}^*(\boldsymbol{u}^\star - \mathcal{S}\bar{\boldsymbol{u}})$, of the $\dot{\gamma}$-based model (figure 3a) and the $k$-based model (figure 3b), projected onto the model space, $\boldsymbol{M}$. The posterior (MAP) data-model discrepancies, $\mathcal{S}^*(\boldsymbol{u}^\star - \mathcal{S}\boldsymbol{u}^\circ)$, of the $\dot{\gamma}$-based and $k$-based models are displayed in figures 3c and 3d respectively. The slices in figures 3a-3d are taken from the mid-plane. The contours are shown in nondimensional units.

### 4.2. *Model parameter learning*

The prior and posterior effective viscosity fields are compared in figures 4 and 5. To enforce the algebraic model (2.9) parameters $\alpha$, $\beta$, $c$ and $\mu_l$, to be positive, we assimilate the logarithms of these parameters and assign Gaussian prior distributions to $\log(\alpha)$, $\log(\beta)$, $\log(c)$ and $\log(\mu_l)$. The prior expected values and standard deviations of the viscosity models are provided in tables 1 and 2. For both models, the inferred mixing lengths deviate more significantly from the prior assumptions, indicating that the posterior is strongly informed by the data. We see that the magnitude of the mixing length near the inlet, and by extension the effective viscosity, is reduced throughout the data assimilation to match the length of the measured jet.

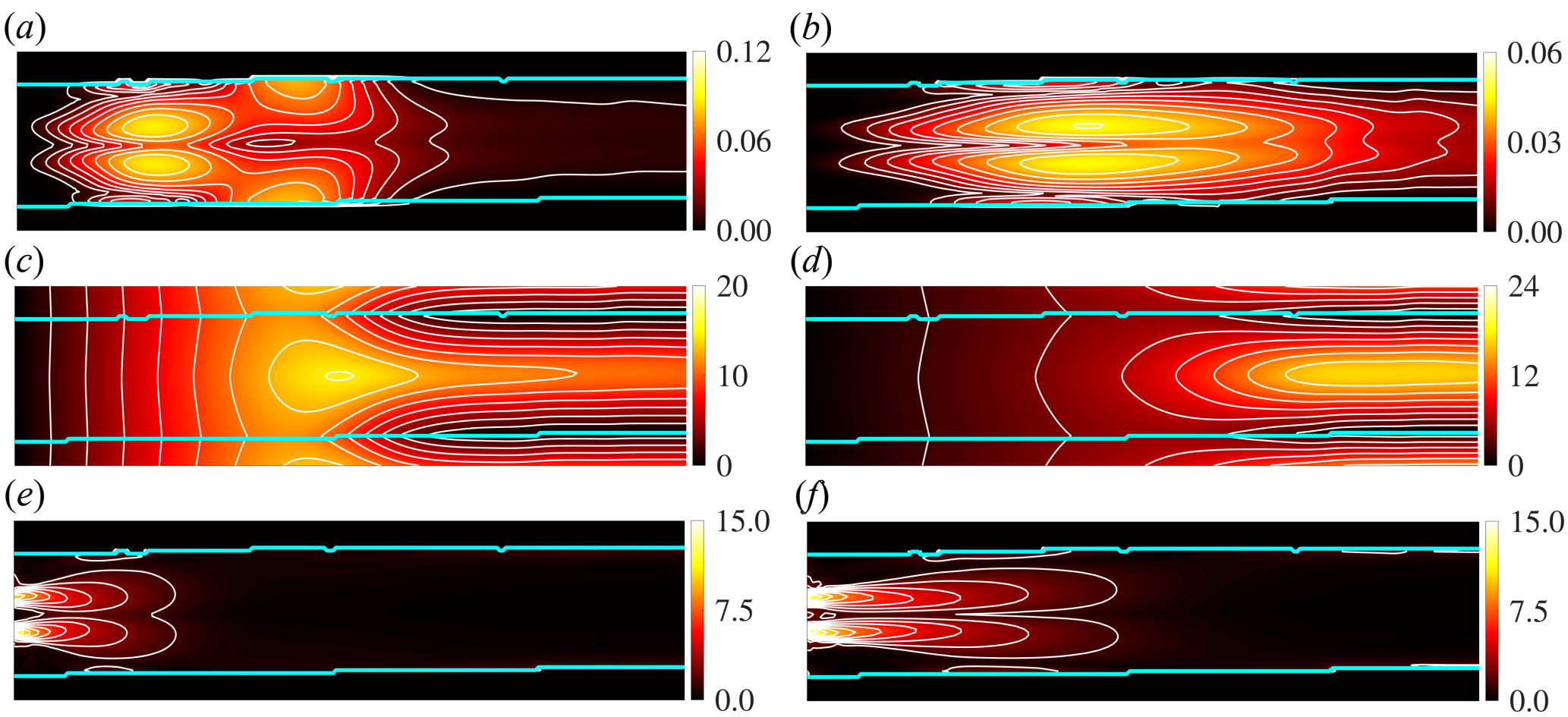

Figure 4. Nondimensional contours of the prior (figure 4a) and posterior (figure 4b) effective viscosity fields predicted by the $\dot{\gamma}$-based model. The prior and posterior mixing lengths are depicted in figure 4c and figure 4d respectively. The corresponding shear rate magnitude profiles are displayed figures 4e and figures 4f. The solid boundaries are indicated by cyan lines.

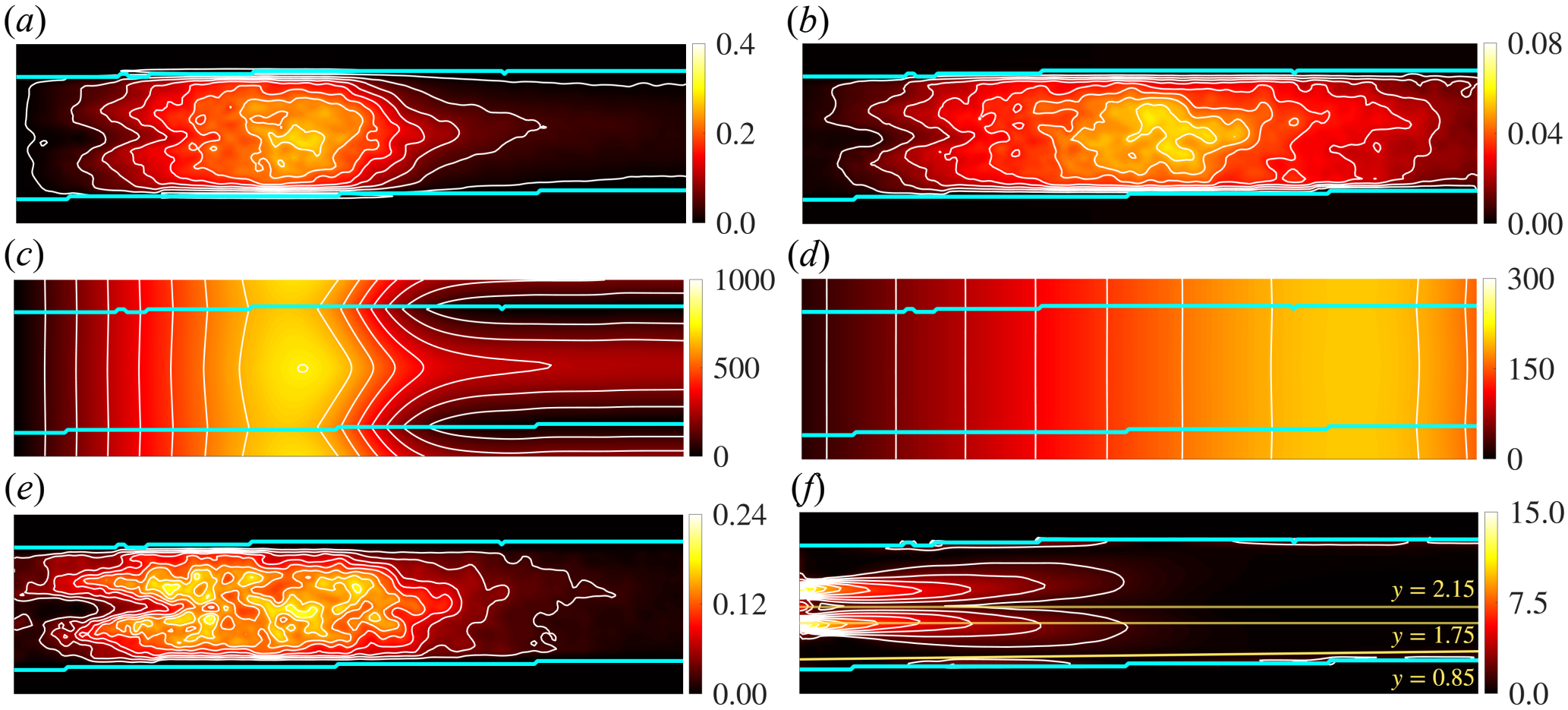

Figure 5. Nondimensional contours of the prior (figure 5a) and posterior (figure 5c) effective viscosity fields predicted by the $k$-based model. The prior and posterior mixing lengths are depicted in figure 5c and figure 5d respectively. The filtered TKE profile is shown in figure 5e. Longitudinal slices at $y = 0.85$, $y = 1.75$ and $y = 2.15$, overlayed onto the posterior shear rate magnitude profile, are shown as yellow solid lines in figure 5f.

Within the jet shear layer, the inferred viscosity levels for the $\dot{\gamma}$-based and $k$-based models agree closely, as can be seen in figure 6a. This is because the rate of entrainment of ambient fluid, driven by turbulent momentum transport in the shear layer, is tightly connected to the width of the jet, for which the data are very informative. In contrast, the effective viscosities deviate more significantly near the walls (figure 6b) and in the jet core (figure 6c), because the local signal-to-noise ratio (SNR) is lower, meaning that the data have low information content in these regions. These discrepancies, paired with the non-overlapping confidence intervals in figure 6, indicate that algebraic models lack the mathematical flexibility to fit flow MRI data across the entire flow domain.

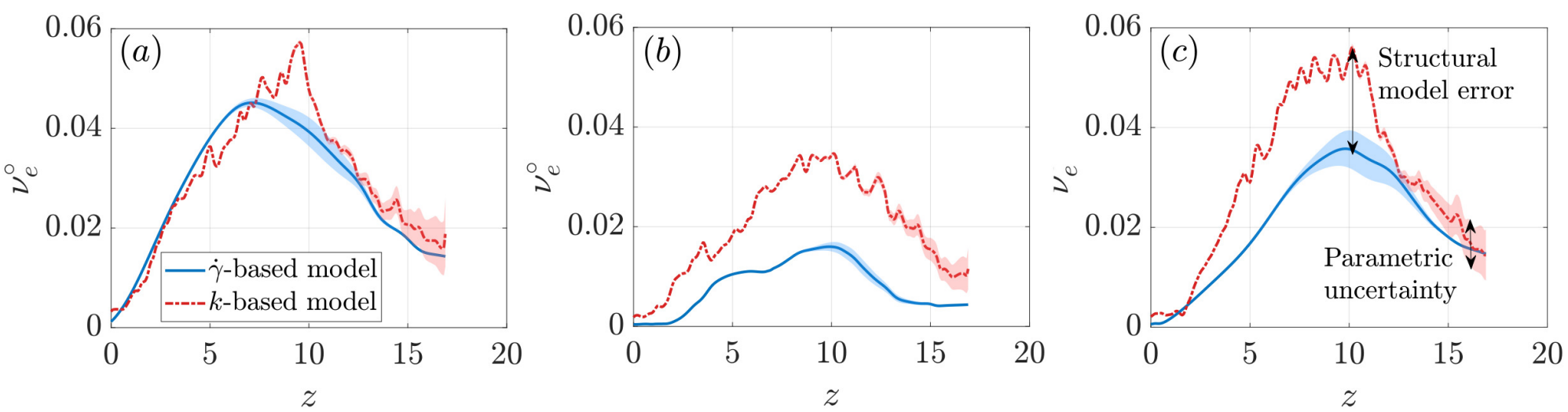

Figure 6. Longitudinal slices of inferred effective viscosity field, $\nu_e^\circ$, taken from the shear layer (figure 6a, $y = 1.75$), the near-wall region (figure 6b, $y = 0.85$), and the jet core (figure 6c, $y = 2.15$). The horizontal and vertical axes are normalised by the reference velocity $U_{\text{ref}} = 1\,\text{m/s}$ and lengthscale $L_{\text{ref}} = 1\,\text{cm}$. The 99% confidence intervals for the $\dot{\gamma}$-based and $k$-based models are shown as blue and red shaded regions, respectively.

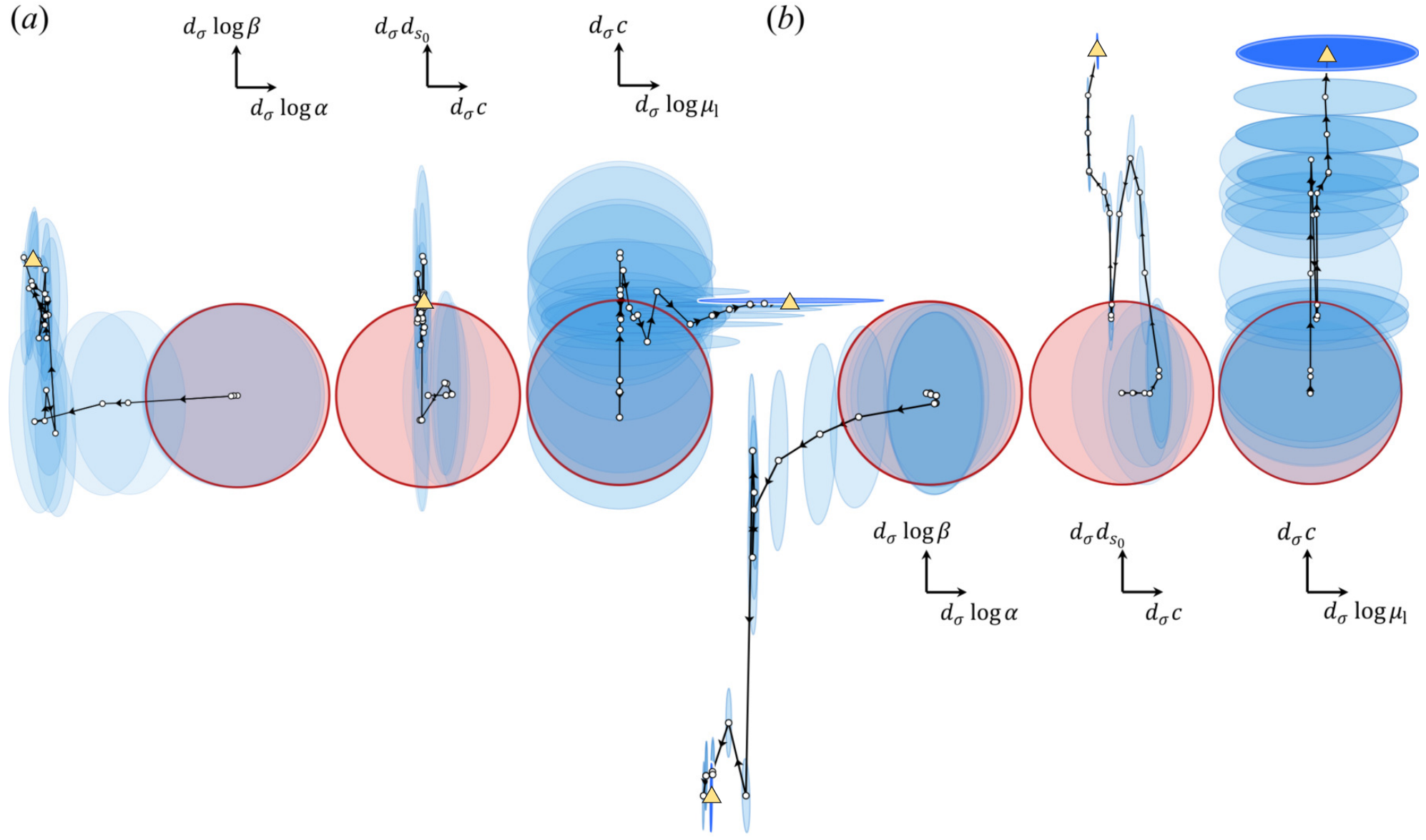

Figure 7. Posterior p.d.f. evolution of the parameters of the $\dot{\gamma}$-based (figure 7a) and $k$-based models (figure 7b). The joint prior and posterior distributions are represented by red circles and blue ellipses, respectively. Each disc corresponds to one standard deviation of posterior uncertainty normalised by the prior uncertainty. The expected values of the model parameters are indicated by small white circles. The MAP points are indicated by yellow triangles. The axes are such that $d_\sigma x := (x - \bar{x})/\sigma_{\bar{x}}$, where $\bar{x}$ is the prior expected value and $\sigma_{\bar{x}}$ is the prior standard deviation.

In figure 7, we plot the evolution of the posterior p.d.f. of the $\dot{\gamma}$-based (figure 7a) and $k$-based (figure 7b) model parameters against their prior distribution parameters. The blue discs represent one standard deviation of posterior uncertainty normalised by the prior uncertainty. The MAP estimates of the means and standard deviations of these parameters are summarised in tables 1 and 2. As the data are assimilated, the discs shrink, indicating a decrease in parametric uncertainty. While both models show strong decreases in parametric uncertainty through data assimilation, the $\dot{\gamma}$-based model achieves a greater reduction than the $k$-based model, probably due to the latter's dependence on TKE data, which is noisy. In both cases, $\alpha$ and $\beta$ move significantly away from their prior expected values. Compared with the mixing length model parameters, there is insufficient information in the measurements to infer $\mu_l$. As a result, the uncertainty in $\mu_l$ reduces little throughout the data assimilation. Further collapse of the uncertainty in $\mu_l$ would require information from a separate rheometry experiment to be provided (Kontogiannis et al. 2025a).

| | | Prior | | |
|---|---|---|---|---|
| $\alpha$ | $\beta$ | $c$ [cm] | $d_{s_0}$ [cm] | $\mu_l$ [mPa.s] |
| $2.00^{+6.96}_{-1.55}$ | $10.00^{+14.60}_{-5.93}$ | $8.50 \pm 7.50$ | $0.00 \pm 15.00$ | $4.20^{+0.13}_{-0.12}$ |
| | | Inferred (MAP estimate) | | |
| $\alpha$ | $\beta$ | $c$ [cm] | $d_{s_0}$ [cm] | $\mu_l$ [mPa.s] |
| $0.66^{+0.01}_{-0.01}$ | $15.60^{+0.56}_{-0.54}$ | $10.86 \pm 0.18$ | $-0.22 \pm 0.05$ | $4.29^{+0.13}_{-0.13}$ |

Table 1. Compound mixing length parameters of the $\dot{\gamma}$-based model. The upper uncertainty bounds for $\alpha$, $\beta$, $c$ and $\mu_l$ differ from the lower bounds because their p.d.f.s are defined on the logarithms of these parameters.

| | | Prior | | |
|---|---|---|---|---|
| $\alpha$ | $\beta$ | $c$ [cm] | $d_{s_0}$ [cm] | $\mu_l$ [mPa.s] |
| $115.58^{+1364.71}_{-106.56}$ | $244.69^{+4670.08}_{-232.51}$ | $8.50 \pm 7.50$ | $0.00 \pm 15.00$ | $4.20^{+0.13}_{-0.12}$ |
| | | Inferred (MAP estimate) | | |
| $\alpha$ | $\beta$ | $c$ [cm] | $d_{s_0}$ [cm] | $\mu_l$ [mPa.s] |
| $15.86^{+0.21}_{-0.21}$ | $3.75^{+7.56}_{-2.51}$ | $17.19 \pm 1.41$ | $-1.11 \pm 0.07$ | $4.27^{+0.13}_{-0.13}$ |

Table 2. As in table 1, but for the $k$-based model.

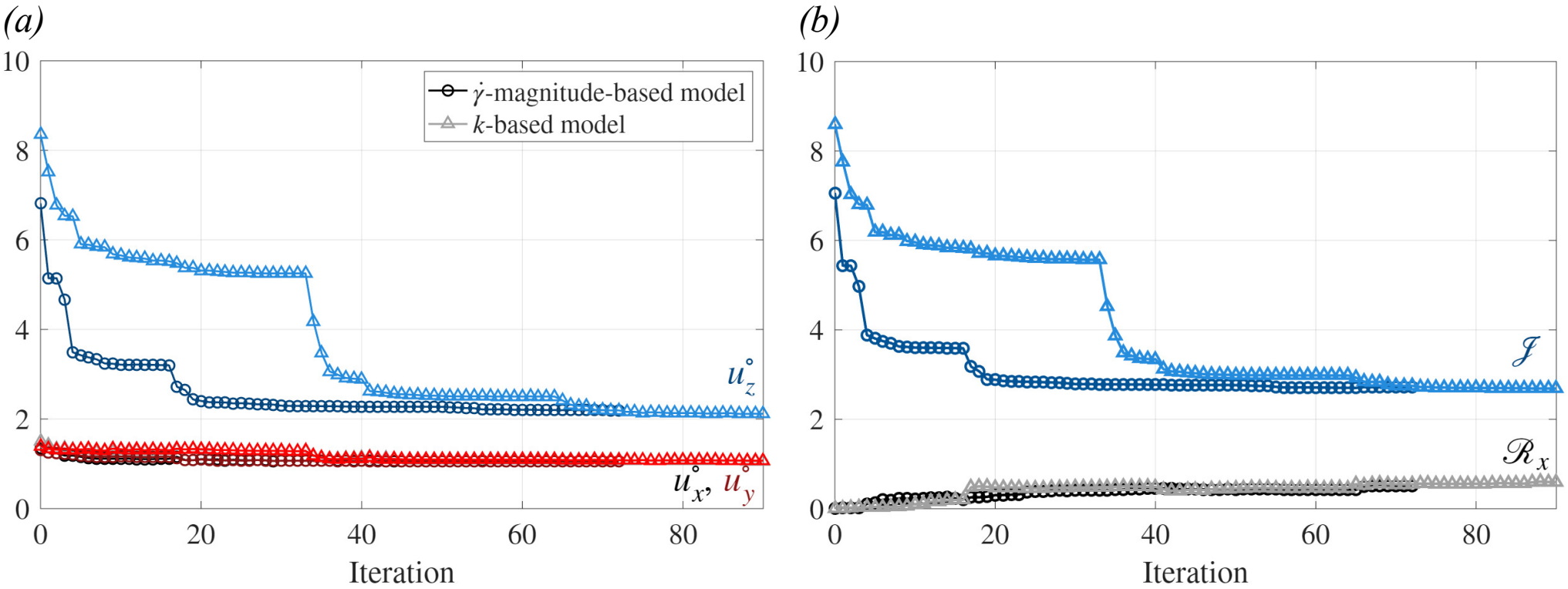

Figure 8. Optimisation log of the data-model discrepancy $\mathscr{E}_{u_i} := \frac{1}{|\Omega|^{1/2}}||u_i^\star - Su_i^\circ||_{C_{u_i}}$ of each velocity component (figure 8a). The objective functional, $\mathscr{J}$, and the prior regularisation term, $\mathscr{R}_x := \frac{1}{2}||\boldsymbol{x} - \bar{\boldsymbol{x}}||_{C_{\bar{x}}}$, are presented in figure 8b.

The convergence history of the data-model discrepancies $\mathscr{E}(u_i) := \frac{1}{|\Omega|^{1/2}}||u_i^\star - Su_i^\circ||_{C_{u_i}}$ and the objective functional $\mathscr{J}$ are shown in figure 8. A discrepancy $\mathscr{E}(u_i) \gg 1$ indicates that the model has not assimilated all the information in the data. Conversely, $\mathscr{E}(u_i) \ll 1$ suggests the inferred models are overfitting the data. The initial decrease of the cost functional is associated with the adjustment of the jet length. We see that 90% error reduction occurs in the first 17 iterations for the $\dot{\gamma}$-based model, and 36 iterations for the $k$-based model. Starting from prior residuals of $(\mathscr{E}(\bar{u}_x), \mathscr{E}(\bar{u}_y), \mathscr{E}(\bar{u}_z)) = (1.34, 1.30, 6.81)$ and $(\mathscr{E}(\bar{u}_x), \mathscr{E}(\bar{u}_y), \mathscr{E}(\bar{u}_z)) = (1.50, 1.39, 8.35)$, the algorithm converges in around 65 iterations to posterior residuals of $(\mathscr{E}(u_x^\circ), \mathscr{E}(u_y^\circ), \mathscr{E}(u_z^\circ)) = (1.07, 1.07, 2.19)$ and

$(\mathscr{E}(u_x^\circ), \mathscr{E}(u_y^\circ), \mathscr{E}(u_z^\circ)) = (1.06, 1.07, 2.09)$ for the $\dot{\gamma}$- and $k$-based models, respectively. While significant reductions in data-model discrepancy are observed, $\mathscr{E}(u_z^\circ)$ remains large across models due to the modelling and systematic errors discussed in section 4.1.

## 5. Summary and conclusions

In this study, we extend the Bayesian inverse Navier–Stokes framework developed by Kontogiannis et al. (2025b) to reconstruct experimental measurements of a confined turbulent jet, while simultaneously approximating the distributions of the most likely parameters and inlet boundary conditions of two algebraic turbulence models. The models, whose parameters have been inferred from the data, are both capable of reconstructing the flow field without overfitting, while also filtering out the observational noise and artefacts. We observe a significant decrease in parameter uncertainty throughout the data assimilation, and find that the models converge in relatively few iterations. To further reduce the posterior data-model discrepancy, the Bayesian inversion framework can be applied to more expressive turbulence models, such as one- or two-equation models. These closures are expected to improve the accuracy of the flow reconstruction and the ability to generalise to unseen flow cases. In particular, models that include a transport equation for $k$ will allow for the assimilation of TKE data as well as velocimetry data, and are the subject of ongoing work.

**Acknowledgements** C.N. gratefully acknowledges financial support from EPSRC and the Harding Distinguished Postgraduate Scholars Programme. A.K. acknowledges financial support from an ESPRC National Fellowship in Fluid Dynamics.

**Declaration of Interests.** The authors report no conflict of interest.